# A NEW MODEL OF FRINGING CAPACITANCE AND ITS APPLICATION TO THE CONTROL OF PARALLEL-PLATE ELECTROSTATIC MICRO ACTUATORS

*Mehran Hosseini, Guchuan Zhu, and Yves-Alain Peter*

École Polytechnique de Montréal
C.P. 6079, Succursale centre-ville
Montréal, QC, Canada H3C 3A7

**ABSTRACT**

Fringing field has to be taken into account in the formulation of electrostatic parallel-plate actuators when the gap separating the electrodes is comparable to the geometrical dimensions of the moving plate. Even in this case, the existing formulations often result in complicated mathematical models from which it is difficult to determine the deflection of the moving plate for given voltages and therefore to predict the necessary applied voltages for actuation control. This work presents a new method for the modeling of fringing field, in which the effect of fringing field is modeled as a serial capacitor. Numerical simulation demonstrates the suitability of this formulation. Based on this model, a robust control scheme is constructed using the theory of input-to-state stabilization (ISS) and back-stepping state feedback design. The stability and the performance of the system using this control scheme are demonstrated through both stability analysis and numerical simulation.

## 1. INTRODUCTION

In the most popular model of electrostatic parallel-plate actuators, only the main electrical field (perpendicular to both electrodes) is considered and the capacitance of the structure is computed by

$$C = \varepsilon WL / G \qquad (1)$$

where $W$ and $L$ are the width and the length of electrodes, respectively, $G$ indicates the separation distance, and $\varepsilon$ is the permittivity in the gap. This formulation leads to a simple model of parallel plate devices, but it is not accurate when the gap size separating the electrodes, is comparable to the geometrical extend of the plates. The capacitance of the structure including the effect of fringing field can be computed by Laplace formula. Although this formula can be used in finite element methods and leads to the most accurate estimation of the real values, it is not susceptible to analytic calculations. Assuming zero thicknesses for the plates, several approximate analytical formulae have been developed for the capacitance in the presence of the fringing field, such as H. B. Palmer [1] and R.S. Elliot models [2]. A number of other formulae have also been recommended assuming finite thicknesses for the electrodes [3-8].

Although the recommended equations consider fringing field in the modeling of parallel-plate capacitors, these formulations provide only approximate expression and are not mathematically simple. In particular, these formulas are highly nonlinear and it is difficult to determine the deflection of the moving plate for a given capacitance. Therefore they are not suitable for predicting the applied voltages for actuation control.

In this work, we present a new method for the modeling of fringing capacitance, in which the effect of fringing field is considered as a serial time-varying capacitor. The value of the serial capacitor and its relationship with deflections is not easy to determine. To overcome this difficulty, a robust control scheme is considered for actuation. With such a control scheme, the knowledge of the relationship between serial capacitance and deflection is not required, but only its variation boundaries.

The proposed formulation can easily cope with other type of modeling errors, such as parallel parasitics and parametric uncertainties, thus it can be applied to the control of more generic MEMS devices.

A robust control scheme is constructed using the theory of input-to-state stabilization (ISS) and back-stepping state feedback design [9-10]. This controller is based on the model of an ideal parallel-plate actuator, and is robust against parasitics and parametric uncertainties, which are considered as disturbances to the system. The stability and the performance of the system using this control scheme are demonstrated through both stability analysis and numerical simulation.

The rest of the paper is organized as follows. Section 2 presents the model of fringing field. In section 3

the dynamics of the driving circuit are established. Section 4 is devoted to the construction of an ISS control law. The simulation results are reported in Section 5 and Section 6 contains concluding remarks.

## 2. MODELING OF FRINGING FIELD

Consider the Palmer formula for the capacitance of a rectangular parallel-plate structure [1]:

$$C_f = \frac{\varepsilon_0}{G} WL \left(1 + \frac{G}{\pi W} + \frac{G}{\pi W} \ln\left(\frac{2\pi W}{G}\right)\right) \times \left(1 + \frac{G}{\pi L} + \frac{G}{\pi L} \ln\left(\frac{2\pi L}{G}\right)\right). \quad (2)$$

As will be seen, this equation conforms to the simulation results to a great extent.

To present the accuracy of different formulations, simulation results of capacitance of a parallel-plate actuator as a function of separation distance, has been compared to the equivalent approximate results using (1) (without fringing field (FF)) and (2) (with FF), given rectangular electrodes ($600\mu m \times 300\mu m$) and initial separation distance $G_0 = 305\mu m$.

As can be seen from Fig_1, at the initial stage when gap is comparable to the dimensions of the electrodes Equation (1) considerably underestimates the real value (almost 0.35 of the real one). Decreasing the gap will attenuate the effect of fringing field and when both *W/G* and *L/G* become more than 100, the difference between the two becomes less than 5%. On the other hand 2-D Palmer formula gives a close approximation to the simulation results (6% error for the initial gap, decreasing with gap reduction).

It can be seen that the effect of fringing field is an increase in the capacitance of the simple device (without FF) and, consequently, the electrostatic force. It has to be noted that (as shown in Fig_1) the extra capacitance due to the fringing field decreases as the gap closes. Therefore, we can use an over-estimated capacitor (which follows the simple model of (1)) combined with an appropriate variable serial capacitor to represent the total capacitance of the device.

Based on this idea, we consider a parallel-plate actuator with the same dimensions mentioned for the analysis of Fig_1 and we develop a model for this device, including a substitute capacitor and a serial capacitor. The substitute capacitor is selected to have the same capacitance as the real device at the initial separation distance between electrodes, but to follow Equation (1) instead, when the gap decreases. Obviously, to satisfy these conditions, the substitute capacitor must have larger electrode areas than the real one. The value of the introduced serial capacitor is time-varying (actually as a function of the gap) and its relationship with deflection is not easy to determine. To overcome this difficulty, a robust control scheme is considered for actuation. With such a control scheme, the knowledge of the relationship between serial capacitance and deflection is not required, but only its variation boundaries.

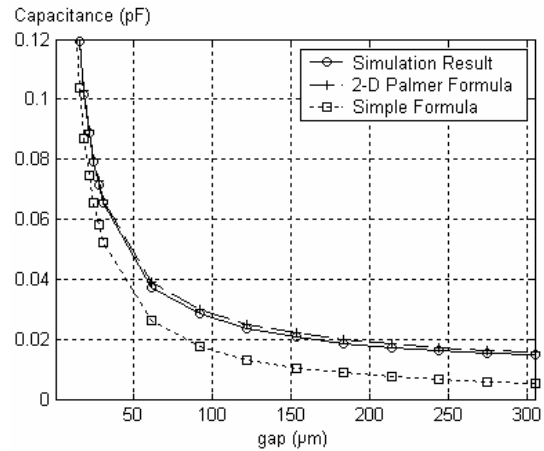

Fig. 1: Comparison of the simulation results with analytical formulae.

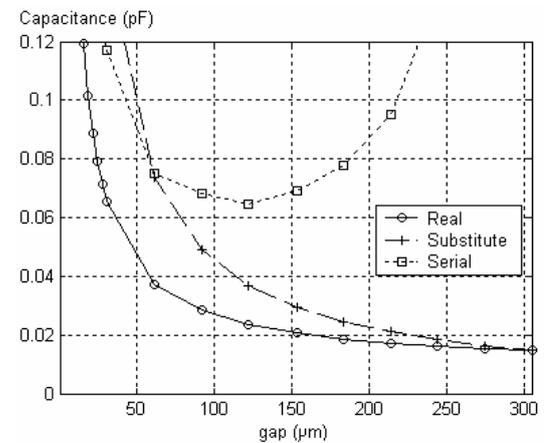

Fig. 2: Values of the real (simulation), its substitute and the serial capacitance.

We use Finite Element Analysis tool (ConventorWare) to determine the boundaries of parasitics. The results are shown in Fig_2. The simulation has been done for a rigid moving electrode, but for the case of a deformable electrode, the procedure will be the same.

The value of the substitute capacitor at the initial gap is equal to $1.474 \times 10^{-2}$ pF (the same as the real capacitor at the same position). Expect for the initial separation distance, there is a difference between the value of substitute capacitor (which follows (1)) and the real one. The role of the serial capacitor is to compensate this difference. As shown in the figure, this serial capacitance is infinite at the initial gap, and as the gap decreases its value becomes closer to the real capacitance. There is a minimum for the value of serial capacitance that is equal to $6.47 \times 10^{-2}$ pF for this structure. We use this boundary value in the control algorithm.

## 3. EQUATION OF MOTION AND DYNAMICS OF DRIVING CIRCUIT

In this work, the parallel parasitic capacitance, due to e.g. the layout, is also considered. The equivalent circuit of the device is shown in Fig. 3, in which $I_S$, $V_S$, and $V_a$ are the source current, the applied voltage, and the actuation voltage, respectively, and $C_{sp}$ and $C_{pp}$ represent the serial and parallel parasitic capacitances, respectively. Denoting by $Q_a$ the charge on the device and by $A$ the plate area, the equation of motion of the parallel-plate actuator is then given by (see e.g. [11]):

$$m\ddot{G} + b\dot{G} + k(G - G_0) = -\frac{Q_a^2}{2\varepsilon A} \quad (3)$$

where $m$, $b$, and $k$ are the mass of the movable upper electrode, the damping coefficient, and the elastic constant, respectively. The actuator is driven by a voltage source.

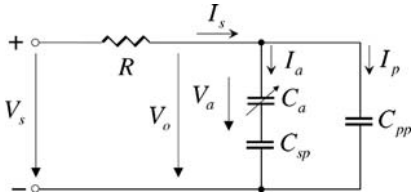

Fig. 3: Equivalent circuit of 1DOF parallel-plate electrostatic actuator with parasitics.

Applying Kirchhoff laws yields:

$$I_s R = V_s - V_a, \quad I_s = I_a + I_p = \dot{Q}_a + \dot{Q}_{pp}.$$

Since the voltage across the actuator is:

$$V_a = \frac{C_{sp}}{C_a + C_{sp}} V_o = \frac{Q_a}{C_a},$$

it follows

$$Q_{pp} = \frac{C_{pp}(C_a + C_{sp})}{C_a C_{sp}} Q_a$$

and

$$\dot{Q}_{pp} = -\frac{C_{pp} Q_a}{C_a^2} \dot{C}_a + \frac{C_{pp}(C_a + C_{sp})}{C_a C_{sp}} \dot{Q}_a.$$

The dynamic equation of the electrical subsystem is therefore given by

$$\dot{Q}_a(t) = \frac{1}{R\left(1 + \rho_p \rho_s + \rho_p \dfrac{G}{G_0}\right)} \times \left(V_s - \left(\frac{G}{\varepsilon A} + \rho_s \frac{G_0}{\varepsilon A} + R\rho_p \frac{\dot{G}}{G_0}\right) Q_a\right), \quad (4)$$

where

$$\rho_p = \frac{C_{pp}}{C_0}, \rho_s = \frac{C_0}{C_{sp}},$$

with $C_0 = \varepsilon A/G_0$, the capacitance of the device at the initial Gap $G_0$.

In our modeling, $\rho_p$ and $\rho_s$ represent the influence of parasitics. When their value is set to zero, the dynamics of the electrical subsystem will be reduced to the one for ideal devices.

It can be seen from (4) that the parallel parasitic capacitance will not change the static behavior of the device. On the other hand, however, the dynamics of the electrical subsystem will be affected: the bigger the parasitic capacitance, the slower the dynamics of the driving circuit. Consequently, the performance of the device will be degraded if the parasitic capacitance is not taken into account in the design of the control system. However, the serial parasitic capacitance will affect both the static and the dynamic behavior of the system. Consequently, in open-loop control schemes, this may result in significant regulation errors. Note that the serial parasitics will move the pull-in position beyond 1/3 of the full gap. This matches the observations reported in the literature on devices in the presence of fringing field (see, e.g., [12]).

To make the system analysis and control design easier, we transform (3) and (4) into normalized coordinates by changing the time scale, $\tau = \omega_0 t$, and performing a normalization as follows [13]:

$$x = 1 - \frac{G}{G_0}, q = \frac{Q_a}{Q_{pi}}, u = \frac{V_s}{V_{pi}}, i = \frac{I_s}{V_{pi} \omega_0 C_0}, r = \omega_0 C_0 R,$$

where $V_{pi} = \sqrt{8kG_0^2/27C_0}$ is the nominal pull-in voltage, $Q_{pi} = \frac{2}{3} C_0 V_{pi}$ the nominal pull-in charge, $\omega_0 = \sqrt{k/m}$ the undamped natural frequency, and $\zeta = b/2m\omega_0$ the damping ratio. We then have

$$\dot{q}(t) = \frac{1}{r\left(1+\rho_p(1-x)+\rho_p\rho_s\right)}\left(\frac{2}{3}u - (1-x)q - \rho_s q + r\rho_p \dot{x}q\right). \quad (5)$$

Let $x_1 = x$, $x_2 = v = \dot{x}$, and $x_3 = q^2$. System (3)-(4) can then be written in the new coordinates as

$$\dot{x}_1 = x_2$$
$$\dot{x}_2 = -2\zeta x_2 - x_1 + \frac{1}{3}x_3 \quad (6)$$
$$\dot{x}_3 = \beta\left(\frac{4\sqrt{x_3}}{3}u - 2(1-x_1)x_3 - 2\rho_s x_3 + 2r\rho_p x_2 x_3\right)$$

where

$$\beta = \frac{1}{r(1+\rho_p(1-x)+\rho_p\rho_s)}, \quad (7)$$

which is a function of position $x$. System (6) is defined on the state space

$$X = \{(x_1, x_2, x_3) \subset R^3 \mid x_1 \in [0,1], x_3 \geq 0\}.$$

Since we will deal only with normalized quantities, we can use $t$ to denote the time and omit the qualifier "normalized."

## 4. CONTROL SYNTHESIS

### 4.1. Preliminaries of Input-to-State Stability

The concept of input-to-state stability is introduced by Sontag in [14] and ISS-based control system design is a well-known tool in the field of system control. We present here only the notations required in the development of the control law. The interested reader is referred to, for example, [9,10] for a formal presentation.

The following comparison functions are required for presenting the method of input-to-state stabilization. A function $\alpha:[0,a) \to [0,\infty)$ belongs to class-$\mathcal{K}$ if it is continuous, strictly increasing, and $\alpha(0) = 0$. If $a = \infty$ and $\alpha$ is unbounded, the function is said to belong to $\mathcal{K}_\infty$. A function $\beta:[0,a)\times[0,\infty) \to [0,\infty)$ is said to belong to class-$\mathcal{KL}$ if it is nondecreasing in it first argument, nonincreasing in its second argument and $\lim_{s\to 0^+}\beta(s,t) = \lim_{t\to\infty}\beta(s,t) = 0$.

The system
$$\dot{x} = f(x,u) \quad (8)$$
is said to be input-to-state stable if for any $x(0)$ and for any input $u(\cdot)$ continuous and bounded on $[0,\infty)$ the solution exists for all $t \geq 0$ and satisfies

$$|x(t)| \leq \beta(x(0),t) + \gamma\left(\sup_{0\leq\tau\leq t}|u(\tau)|\right), \forall t \geq 0, \quad (9)$$

where $\beta(s,t) \in \mathcal{KL}$ and $\gamma(s) \in \mathcal{K}$.

A useful property of ISS systems is that for a cascade of two systems

$$\dot{x}_1 = f_1(x_1, x_2, u)$$
$$\dot{x}_2 = f_2(x_2, u) \quad (10)$$

if the $x_1$-subsystem is ISS with respect to $x_2$ and $u$, and the $x_2$-subsystem is ISS with respect to $u$, then the cascade system is also ISS [9,10].

Note that the method of ISS applied to robust system control by considering disturbances as the inputs to the closed-loop system.

### 4.2. Control Synthesis

In this work, we consider both the parasitics and parametric uncertainties, such as the variations of damping coefficient and loop resistance. We make then the following assumptions on the uncertainties in System (6).

**Assumption 1** *The parasitic capacitances are bounded by known constants:*

$$0 \leq \rho_p \leq \overline{\rho}_p, \; 0 \leq \rho_s \leq \overline{\rho}_s. \quad (11)$$

**Assumption 2** *The damping ratio is positive and bounded and can be written as:*

$$\zeta = \zeta_0 + \Delta\zeta \quad (12)$$

where $\zeta_0$ is positive-valued representing the nominal damping ratio and $\Delta\zeta$ the modeling error.

**Assumption 3** *The upper and lower bounds for the resistance r are known:*

$$0 < \underline{r} \leq r \leq \overline{r}. \quad (13)$$

Since $x_1 \in [0,1]$, $\beta$ in (7) is bounded as follows:

$$0 < \underline{\beta} \leq \beta \leq \overline{\beta}, \quad (14)$$

where

$$\underline{\beta} = \frac{1}{\overline{r}(1+\overline{\rho}_p(1+\overline{\rho}_s))}, \; \overline{\beta} = \frac{1}{\underline{r}}.$$

We denote furthermore

$$\Delta\beta = \beta - \beta_0, \quad (15)$$

where $\beta_0$ is the nominal value of $\beta$.

In this work, we will consider the tracking problem with $y = x_1$ as the output. Following a classical approach, we choose a sufficiently smooth reference trajectory $y_r$ for $x_1$ as a function of time. Then we will find a controller that makes this trajectory attractive.

A recursive procedure, called also backstepping design (see, e.g., [10] for a detailed presentation), is used in the design of the control law, which results in:

$$x_{2d} = \dot{y}_r - k_1 z_1, \quad (16)$$

$$x_{3d} = 3\left(2\zeta_0 x_2 + x_1 + \dot{x}_{2d} - \kappa_2 \zeta_0 | x_2 | z_2 - k_2 z_2\right), \quad (17)$$

$$u = \frac{3}{4\sqrt{x_3}}\left(2x_3(1-x_1) + \frac{1}{\underline{\beta}}\left((3ab_1 + 3b_2 - k_3 z_3)\right.\right.$$
$$- \kappa_{31}\zeta_0 | b_1 x_2 | z_3 - \kappa_{32}(| ab_1 + b_2 |) z_3 \quad (18)$$
$$\left.\left. - \kappa_{33}\overline{r}\overline{\rho}_p | x_2 | x_3 z_3 - \kappa_{34}\overline{\rho}_s x_3 z_3 \right)\right),$$

where $z_1 = x_1 - y_r$, $z_2 = x_2 - x_{2d}$, $z_3 = x_3 - x_{3d}$ are closed-loop tracking errors, $k_1, k_2, k_3, \kappa_2, \kappa_{31}, \kappa_{32}, \kappa_{33}$, and $\kappa_{34}$ are controller gains, and

$$a = -2\zeta_0 x_2 - x_1 + \frac{1}{3}x_3,$$
$$b_1 = 2\zeta_0 - k_1 - k_2 - \kappa_2 \zeta_0 \left(\text{sgn}(x_2) z_2 + | x_2 |\right),$$
$$b_2 = y_r^{(3)} + k_1 \ddot{y}_r + \left(\kappa_2 \zeta_0 | x_2 | + k_2\right) \dot{x}_{2d} + x_2.$$

It can be shown that the closed-loop error dynamics satisfy

$$| z_1(t) | \leq | z_1(0) | e^{-\frac{1}{2}k_1 t}, \quad (19)$$

$$| z_2(t) | \leq | z_2(0) | e^{-\frac{1}{2}k_2 t} + \sup_{0 \leq \tau \leq t} \mu_2(\tau), \quad \forall t \geq 0, \quad (20)$$

where $\mu_2$ is the following continuous and uniformly bounded function:

$$\mu_2 = \frac{2 | \Delta\zeta x_2 | + | z_3 | / 3}{k_2/2 + \kappa_2 \zeta_0 | x_2 |}, \quad (21)$$

and

$$| z_3(t) | \leq | z_3(0) | e^{-\frac{1}{2}k_3 t} + \sup_{0 \leq \tau \leq t} \mu_3(\tau), \quad \forall t \geq 0, \quad (22)$$

where $\mu_3$ is the following continuous and uniformly bounded function:

$$\mu_3 = \mu_{31} + \mu_{32} + \mu_{33} + \mu_{34} \quad (23)$$

with

$$\mu_{31} = \frac{6 | \Delta\zeta b_1 x_2 |}{\frac{k_3}{8} + \frac{\beta}{\underline{\beta}}\kappa_{31}\zeta_0 | b_1 x_2 |}, \quad \mu_{32} = \frac{3\frac{\Delta\beta}{\underline{\beta}}(| ab_1 + b_2 |)}{\frac{k_3}{8} + \frac{\beta}{\underline{\beta}}\kappa_{32}(| ab_1 + b_2 |)},$$

$$\mu_{33} = \frac{\frac{2\rho_p}{1+\rho_p(1+\rho_s)} | x_2 | x_3}{\frac{k_3}{8} + \frac{\beta}{\underline{\beta}}\kappa_{33}\overline{r}\,\overline{\rho}_p | x_2 | x_3}, \quad \mu_{34} = \frac{\frac{2\rho_s}{r(1+\rho_p(1+\rho_s))} x_3}{\frac{k_3}{8} + \frac{\beta\overline{\rho}_s}{\underline{\beta}}\kappa_{34} x_3}.$$

Therefore, the closed-loop error dynamics of (19), (20), and (22) are all ISS if all the conditions declared in Assumption 1-3 hold, $k_1 > 0$, $k_2 > 0, k_3 > 0, \kappa_2 > 0$, $\kappa_{31} > 0$, $\kappa_{32} > 0$, $\kappa_{33} > 0$, and $\kappa_{34} > 0$. The stability of the overall system is then deduced from the property of cascade interconnected ISS systems. Furthermore, the ultimate bound for the tracking error $z_1$ can be rendered arbitrarily small by choosing the feedback gains $k_1$, $k_2$, and $k_3$, and the damping gain $\kappa_2$, $\kappa_{31}$, $\kappa_{32}$, $\kappa_{33}$, and $\kappa_{34}$ large enough. The construction of the controller and the stability analysis follow the ideas presented in [15,16].

Note that the control $u$ is singular when $x_3 = 0$. This is due to the uncontrollability of System (6) at the initial position. However this situation happens only when the system is at the initial position. It is easy to see that System (6) is stabilizing at the origin with an input $u = 0$. Hence this control will stabilize the system (6) in all air gap except at the origin, for that a stabilizing control is $u = 0$.

### 4.3. Reference Trajectory Design

In general, reference trajectories can be expressed by any sufficiently smooth function $t \mapsto y(t)$, connecting the initial point at time $t_i$ and a desired point at time $t_f$, such that the initial and final conditions are verified. The reference trajectory used in our control schemes is a polynomial of the following form:

$$y_r(t) = y(t_i) + (y(t_f) - y(t_i))\tau^5(t)\sum_{i=0}^{4} a_i \tau^i(t), \quad (24)$$

where $\tau(t) = (t - t_i)/(t_f - t_i)$. For a set-point control, the coefficients in (24) can be determined by imposing the initial and final conditions

$$\dot{y}(t_i) = \dot{y}(t_f) = \ddot{y}(t_i) = \ddot{y}(t_f) = y^{(3)}(t_i) = y^{(3)}(t_f) = 0,$$

and we obtain $a_0 = 126$, $a_1 = -420$, $a_2 = 540$, $a_3 = -315$, and $a_4 = 70$.

The polynomial in (24) is one of the most used reference trajectories in flatness-based control. A more general formulation can be found in [17].

## 5. EXAMPLES AND SIMULATION RESULTS

In our simulation study, the parameters of the nominal plant are $\zeta_0 = 1$, $r_0 = 1$, $\rho_p = 0$, and $\rho_s = 0$. The actuator is supposed to be driven by a bipolar voltage source. Based on the simulation in Section 2 we have $\bar{\rho}_s = 0.226$. Other parameter variations are $\Delta\zeta = 2$, $\bar{r} = 2$, $\bar{\rho}_p = 2$. The simulation results for set-point control of 20%, 40%, 60%, 80%, and 100% deflections are shown in Fig. 4. It can be seen that the proposed control scheme performs well for important parametric uncertainties and parasitic variations.

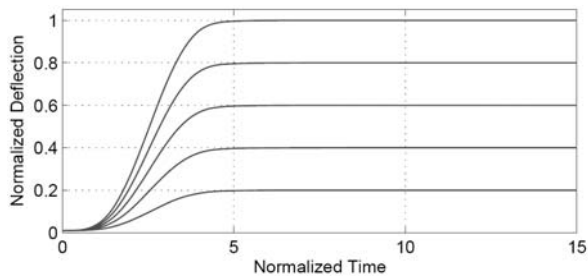

Fig. 4: Simulation results of set-point control.

## 6. CONCLUSIONS

This paper presents a new model in which the effect of fringing field is represented by a time-varying serial capacitor. The exact value of the serial capacitance is not required, but only its boundary represented by the ratio of its minimal value and the equivalent nominal capacitance at the initial position. A Finite Element Analysis tool (ConventorWare) was used to compute the variation range of this serial capacitor. The dynamics of the driving circuits, which cope also with parallel parasitics have then been deduced. A robust control scheme is proposed. Numerical simulations show that the proposed control system has satisfactory performance and robustness vis-à-vis parasitics and parametric uncertainties.

## REFERENCES

[1] B. Palmer, "Capacitance of a parallel-plate capacitor by the Schwartz-Christoffel transformation," *Trans. AIEE*, Vol. 56, pp. 363, March 1927.

[2] R. S. Elliott, *Electromagnetics,* New York, McGraw-Hill, 1966.

[3] W. H. Chang, "Analytic IC-metal-line capacitance formulas," *IEEE Trans. Microwave Theory Tech.*, Vol. MTT-24, pp. 608-611, 1976; also vol. MTT-25, p. 712, 1977.

[4] C. P. Yuan and T. N. Trick, "A simple formula for the estimation of the capacitance of two-dimensional interconnects in VLSI circuits," *IEEE Electron Device Lett.*, Vol. EDL-3, pp. 391-393, 1982.

[5] T. Sakurai and K. Tamaru, "Simple formulas for two- and three-dimensional capacitances," *IEEE Trans. Electron Devices*, Vol. ED-30, pp. 183-185, 1983.

[6] N. Van de Meijs, and J. T. Fokkema, "VLSI circuit reconstruction from mask topology," *Integration*, Vol. 2, pp. 85-119, 1984.

[7] V. Leus and D. Elata, *Fringing Field Effect in Electrostatic Actuators*. Technical report ETR-2004-02, Israel Institute of Technology, 2004.

[8] G. J. Sloggett, N. G. Barton, and S. J. Spencer, "Fringing fields in disc capacitors," *J. Phys., A: Math. Gen.*, vol. 19, pp. 2725–2736, 1986.

[9] E. Sontag, "The ISS philosophy as a unifying framework for stability-like behavior," in *Nonlinear Control in the Year 2000* (Volume 2), A. Isidori, F. Lamnabhi-Lagarrigue, and W. Respondek, Eds. Berlin, Springer-Verlag, 2000, pp. 443–468.

[10] M. Krstić, I. Kanellakopoulos, and P. Kokotović, *Nonlinear and Adaptative Control Design,* John Wiley & Sons Ltd, New York, 1995.

[11] S. Senturia, *Microsystem Desig,*. Kluwer Academic Publishers, Norwell, 2002.

[12] J. Cheng, J. Zhe, and X. Wu, "Analytical and finite element model pull-in study of rigid and deformable electrostatic microactuators," *J. Micromech. Microeng.*, vol. 14, pp. 57–68, 2004.

[13] J. Pont-Nin, A. Rodríguez, and L. Castacer, "Voltage and pull-in time in current drive of electrostatic actuators," *J. Microelectromech. Syst.*, vol. 11, no. 3, pp. 196–205, 2002.

[14] E. Sontag, "Smooth stabilization implies coprime factorization," *IEEE Trans. Automat. Contr.*, vol. 34, pp. 435–443, 1989.

[15] Z.-J. Yang and M. Minashima, "Robust nonlinear control of a feedback linearizable voltage-controlled magnetic levitation system," *T.IEE Japan, vol. 121-C*, no. 7, pp. 1203–1211, 2001.

[16] G. Zhu, J. Penet, and L. Saydy, "Robust control of an electrostatic actuated MEMS in the presence of parasitics and parameter uncertainties," *To appear in the proceeding of the 2006 American Control Conference.*

[17] J. Lévine, *Analyse et Commande des Systèmes Non Linéaires.* [Online] Available: http://cas.ensmp.fr/~levine/ Enseignement/CoursENPC.pdf, 2004.